\documentclass[aps,showpacs,preprint,superscriptaddress]{revtex4}
\usepackage{graphicx}
\usepackage{subfigure}
\usepackage{float}
\usepackage{amsmath}
\usepackage{amsfonts}
\usepackage{color}
\usepackage{bm}
\usepackage{bbm}
\usepackage{txfonts}
\usepackage{array}
\usepackage{amssymb}

\begin{document}

\title{Pair production in inhomogeneous electric fields with phase modulation}
\author{Li-Na Hu}
\affiliation{Key Laboratory of Beam Technology of the Ministry of Education, and College of Nuclear Science and Technology, Beijing Normal University, Beijing 100875, China}
\author{Orkash Amat}
\affiliation{Key Laboratory of Beam Technology of the Ministry of Education, and College of Nuclear Science and Technology, Beijing Normal University, Beijing 100875, China}
\author{Lie-Juan Li}
\affiliation{Key Laboratory of Beam Technology of the Ministry of Education, and College of Nuclear Science and Technology, Beijing Normal University, Beijing 100875, China}
\author{Melike Mohamedsedik}
\affiliation{Key Laboratory of Beam Technology of the Ministry of Education, and College of Nuclear Science and Technology, Beijing Normal University, Beijing 100875, China}
%\author{Adiljan Sawut}
%\affiliation{Key Laboratory for GeoMechanics and Deep Underground Engineering, China University of Mining and Technology, Beijing 100083, China}
\author{B. S. Xie \footnote{bsxie@bnu.edu.cn}}
\affiliation{Key Laboratory of Beam Technology of the Ministry of Education, and College of Nuclear Science and Technology, Beijing Normal University, Beijing 100875, China}
\affiliation{Institute of Radiation Technology, Beijing Academy of Science and Technology, Beijing 100875, China}
\date{\today}

\begin{abstract}
Electron-positron pair production is investigated in spatial inhomogeneous electric fields with high or/and low central frequency as well as sinusoidal phase modulation. It is found that the momentum spectrum (the reduced particle number) is more sensitive to the modulated amplitude (modulated frequency) of the phase. The stronger the modulation parameters are applied, the more remarkable the interference effect in momentum spectrum occur. In particular, for high central frequency field, an extremely good symmetry in momentum spectrum is found while it is destroyed severely when the modulated amplitude becomes large. The reduced particle number can be also enhanced greatly at about a few times or/and one order by the modulation parameters. Moreover, the effect of spatial scales on the reduced particle number are examined carefully and found that it increases rapidly at small spatial scales, while it tends to be a constant at large spatial scales. Two interesting features are revealed for the reduced particle number, i.e., the optimal modulation parameters are found and the same particle number can be got through different set of modulation parameters. The latter findings is important because one can choose different ways of phase modulation to realize the required pair number even if for the optimal pair production.
\end{abstract}
\maketitle

\section{Introduction}

Electron-positron ($e^{-}e^{+}$) pair production from vacuum in strong background fields, the Sauter-Schwinger effect, is one of the well-known nonperturbative predictions of quantum electrodynamics (QED) \cite{Sauter:1931zz,Heisenberg:1935qt,DiPiazza:2011tq}. It has not been verified experimentally due to the current laser intensity $\sim10^{22}~\rm \mathrm{W} / \rm \mathrm{cm}^{2}$ is far less than the critical laser intensity $\sim10^{29}~\rm \mathrm{W} / \rm \mathrm{cm}^{2}$ (the corresponding critical electric field strength is $E_{\mathrm{cr}}=m_e^2c^{3} / e \hbar \approx 1.3 \times 10 ^{16}~\rm {\mathrm{V}/\mathrm{cm}}$, where $m_{e}$ denotes the electron mass and $e$ is magnitude of electron charge) \cite{Schwinger:1951nm,Dunne:2008kc}. Multiphoton pair production is another important mechanism for $e^{-}e^{+}$ pair creation, which has been detected in a laboratory \cite{Burke:1997ew}. Moreover, the pair production can be observed even in the laser field with intensities one or two orders of magnitude lower than the critical value due to the proposed dynamically assisted Schwinger mechanism that combines two laser fields with a low frequency strong field and a high frequency weak field \cite{Schutzhold:2008pz,Bulanov:2010ei}. Fortunately, with the advance of the high-intensity laser technology, the Extreme Light Infrastructure (ELI) \cite{Ringwald:2001ib} and the x-ray free electron laser (XFEL) may achieve subcritical laser intensity, which greatly improves the hope to observe the pair production in the laboratory.

Theoretically, several field configurations have been applied to the study of vacuum pair production, such as the alternating electric field with $N$-pulse \cite{Li:2014psw}, time delay electric field \cite{Li:2017qwd}, the combination of cosine with Gaussian or super-Gaussian pulse external field \cite{Hebenstreit:2009km,Nuriman:2012hn,Dumlu:2011rr}, and so on. Recent studies suggest that the fields with frequency chirp are crucial to the $e^{-}e^{+}$ pair production. It can not only enhance significantly the total particle number but also realize experimental verification by applying the chirped pulse amplification (CPA) technique \cite{Strickland:1985gxr}. At present, the asymmetrical \cite{Dumlu:2010vv,Olugh:2018seh,Ababekri:2020,Li:2021vjf} and symmetrical \cite{Wang,Mohamedsedik:2021pzb} frequency chirp have been studied on pair production in both of spatially homogeneous and inhomogeneous fields. Moreover, the sinusoidal frequency modulation has been used to the investigation of pair production only in homogeneous electric field, and indicated that the momentum distribution and the number density of created particles are sensitive to modulation parameters \cite{Gong:2019sbw}.

On the other hand, the previous investigations show that the spatial inhomogeneity of external fields plays an important role on $e^{-}e^{+}$ pair production, which has displayed some novel features \cite{Kohlf2020Effect,Ababekri:2019dkl,Hebenstreit:2011wk,Kohlfurst:2017hbd,Kohlfurst:2017git}. For example, the self-bunching effect of particles is identified in Schwinger pair production under the electric field with finite spatial scales \cite{Hebenstreit:2011wk}. The ponderomotive force effect is reported in multiphoton process for the small spatial scales of oscillating field \cite{Kohlfurst:2017hbd}. The spin-field interaction is found in Schwinger pair production for spatially inhomogeneous external fields \cite{Kohlfurst:2017git}. However, the sinusoidal frequency modulation has not been considered on $e^{-}e^{+}$ pair production in spatially inhomogeneous fields.

In this work, we investigate the $e^{-}e^{+}$ pair production in spatially inhomogeneous electric fields with sinusoidal phase modulation by using the real-time Dirac-Heisenberg-Wigner (DHW) formalism. The momentum spectrum and the reduced particle number are studied in both high- and low-frequency fields and are found depending strongly on the amplitude and frequency of the modulated phase. It is found that the interference effect and symmetry of momentum spectrum change obviously when the different modulation parameters are applied. The reduced particle number can be also enhanced significantly by the modulated amplitude, while it has different variation by the modulated frequency for different spatial scales. It is evident that in the case of high frequency field, comparable to the small scale where the enhanced particle number is got by the large modulated frequency, however, at the large scale, it is got by the small frequency. Moreover, when modulation parameters are fixed, the effect of spatial scale on the reduced particle number are studied. We obtain the optimal modulation parameters for the particle number in high frequency field and find that the same particle number can be got through different set of modulation parameters. Finally, the corresponding results that we obtained are discussed qualitatively by the semiclassical Wentzel-Kramers-Brillouin (WKB) approach \cite{Brezin1970,Dumlu1011,Strobel2015} and the view point from the action of worldline instanton \cite{Gies:2005bz,Linder:2015}. Note that the natural units $\hbar=c=1$ are applied and all quantities are presented in terms of the electron mass $m$. For example, the spatial and temporal scales of the electric field are in units of $1/m$, and the field frequency is in units of $m$.

The paper is structured as follows. In Sec. \ref{fields}, we review the DHW formalism and the semiclassical WKB approximation method briefly, and introduce the background field to be considered in our work. In Sec. \ref{results}, we show the numerical results for high frequency field with different modulation parameters. In Sec. \ref{result2} the numerical results of the low frequency field with different modulation parameters are presented. Sec. \ref{result3} is discussion. In Sec. \ref{conclusion}, we give a brief conclusion and outlook.
\hspace{-1cm}
\section{Theoretical formalism and field modle}\label{fields}
\hspace{-1cm}
\subsection{The DHW formalism}\label{method}

The DHW formalism is a relativistic phase-space method that has been widely used to investigate vacuum pair creation within arbitrary electromagnetic fields. Since the complete derivation of DHW formalism has been obtained in Refs. \cite{Hebenstreit:2011wk, Kohlfurst:2015zxi, Ababekri:2019qiw}, we only present key points of this method.

We start with the gauge-covariant density operator which is composed of two commutative Dirac field operators, i.e.,
\begin{equation}\label{DensityOperator}
 \hat {\mathcal C}_{\alpha \beta} \left(r , s \right) = \mathcal U \left(A,r,s
\right) \ \left[ \bar \psi_\beta \left( r - s/2 \right), \psi_\alpha \left( r +
s/2 \right) \right],
\end{equation}
where $r=(r_{1}+r_{2})/2$ denotes the center-of-mass coordinate and $s=s_{1}-s_{2}$ represents the relative coordinate. The Wilson line factor
\begin{equation}\label{Wilson line factor}
\mathcal U \left(A,r,s \right) = \exp \left( \mathrm{i} \ e \ s \int_{-1/2}^{1/2} d
\xi \ A \left(r+ \xi s \right)  \right),
\end{equation}
can be used to guarantee the invariant of the density operator. The covariant Wigner operator is obtained via the Fourier transform of Eq. \eqref{DensityOperator}
\begin{equation}\label{WignerOperator}
\hat{\mathcal W}_{\alpha \beta} \left( r , p \right) = \frac{1}{2} \int d^4 s \
\mathrm{e}^{\mathrm{i} ps} \  \hat{\mathcal C}_{\alpha \beta} \left( r , s
\right).
\end{equation}
We can define the covariant Wigner function by taking the vacuum expectation value of Eq. \eqref{WignerOperator} as
\begin{equation}\label{Wigner function}
 \mathbbm{W} \left( r,p \right) = \langle \Phi \vert \hat{\mathcal W} \left( r,p
\right) \vert \Phi \rangle.
\end{equation}
Being a Dirac-matrix valued quantity, the Wigner function can be expanded in terms of $16$ covariant Wigner coefficients
\begin{equation}\label{decomposed}
\mathbbm{W} = \frac{1}{4} \left( \mathbbm{1} \mathbbm{S} + \textrm{i} \gamma_5
\mathbbm{P} + \gamma^{\mu} \mathbbm{V}_{\mu} + \gamma^{\mu} \gamma_5
\mathbbm{A}_{\mu} + \sigma^{\mu \nu} \mathbbm{T}_{\mu \nu} \right).\
\end{equation}
Since we are dealing with the $e^{-}e^{+}$ pair production, and we want to describe it as an initial value problem, the equal-time Wigner function can be obtained by taking the energy average of the covariant Wigner function
\begin{align}
 \mathbbm{w} \left( \mathbf{x}, \mathbf{p}, t \right) = \int \frac{d p_0}{2 \pi}
\ \mathbbm{W} \left( r,p \right).
\end{align}

When the $e^{-}e^{+}$ pair production in time dependent field with spatial inhomogeneity of $x$-axis ($1+1$ dimensional time-space) is investigated, the complete DHW equations of motion can be reduced to the following form
\begin{align}
 &D_t \mathbbm{s} - 2 p_x \mathbbm{p} = 0 , \label{pde:1}\\
 &D_t \mathbbm{v}_{0} + \partial _{x} \mathbbm{v}_{1} = 0 , \label{pde:2}\\
 &D_t \mathbbm{v}_{1} + \partial _{x} \mathbbm{v}_{0} = -2 m \mathbbm{p} , \label{pde:3}\\
 &D_t \mathbbm{p} + 2 p_x \mathbbm{s} = 2 m \mathbbm{v}_{1} , \label{pde:4}
\end{align}
with the pseudodifferential operator
\begin{equation}\label{pseudoDiff}
 D_t = \partial_{t} + e \int_{-1/2}^{1/2} d \xi \,\,\, E_{x} \left( x + i \xi \partial_{p_{x}} \, , t \right) \partial_{p_{x}}.
\end{equation}
The corresponding vacuum initial conditions are
\begin{equation}\label{vacuum-initial}
{\mathbbm{w}}_{0 \, \rm{vac}} = - \frac{2m}{\Omega} \, ,
\quad  {\mathbbm{w}}_{2 \, \rm{vac}} = - \frac{2{ p_x} }{\Omega} \,  ,
\end{equation}
where $\Omega$ represents the one-particle energy, which can be expressed as $\Omega=\sqrt{p_{x}^{2}+m^2}$. By subtracting these vacuum terms, the modified Wigner component can be written as
\begin{equation}
\mathbbm{w}_{k}^{v} \left( x , p_{x} , t \right) = \mathbbm{w}_{k} \left( x , p_{x} , t \right) - \mathbbm{w}_{{k} \, \rm{vac}}\left(p_{x} \right),
\end{equation}
here $\mathbbm{w}_{k}$ is the Wigner component in Eqs. \eqref{pde:1}-\eqref{pde:4}, and we define $\mathbbm{w}_{0} = \mathbbm{s}$, $\mathbbm{w}_{1} = \mathbbm{v}_{0}$, $\mathbbm{w}_{2} = \mathbbm{v}_{1}$ and $\mathbbm{w}_{3} = \mathbbm{p}$. The $\mathbbm{w}_{{k} \, \rm{vac}}$ denotes the corresponding vacuum initial condition in Eq. \eqref{vacuum-initial}. The particle number density can be expressed as
\begin{equation}\label{particle number density}
n \left( x , p_{x} , t \right) = \frac{m  \mathbbm{s}^{v} \left( x , p_{x} , t \right) + p_{x}  \mathbbm{v}_{1}^{v} \left( x , p_{x} , t \right)}{\Omega \left( p_{x} \right)}.
\end{equation}
We can obtain the particle number density of momentum space via integrating Eq. \eqref{particle number density} with respect to $x$,
\begin{equation}\label{momentum distribution}
n\left( p_x,t \right) = \int dx \, n \left( x , p_{x} , t \right).
\end{equation}
Consequently, the total particle yield of the whole phase space can be written as
\begin{equation}\label{Num}
N\left(t \right) = \int dx\,dp_x n \left( x , p_{x} , t \right).
\end{equation}

Moreover, in order to extract the nontrivial effect of spatial scale $\lambda$, we calculate the reduced quantities $\bar{n}\left( p_{x}, t \right)\equiv n\left( p_{x}, t \right)/\lambda$ and $\bar{N}\left(t\rightarrow\infty\right)\equiv N\left(t\rightarrow\infty\right)/{\lambda}$.

\subsection{Semiclassical WKB approximation}\label{results}

To further discuss the momentum spectrum and the particle number of the created particles, the semiclassical WKB approximation method is introduced. The pair production from vacuum is similar to the over-the-barrier scattering problem in quantum mechanics, meanwhile, the physical picture can be reflected in the typical turning point structure. The approximate expression of the particle creation rate can be described as
\begin{equation}\label{AE}
\begin{aligned}
N\approx \sum\limits_{t_{i}}e^{-2K_{i}}- \sum\limits_{t_{i}\neq t_{j}} 2\cos(2\theta_{(i,j)})
e^{-K_{i}-K_{j}},
\end{aligned}
\end{equation}
with
$$ K_{i}=\left\vert \int_{t_{i}^{*}}^{t_{i}}{\omega_{\bm p}(t)dt} \right\vert,$$
and
$$\theta_{(i,j)}=\int_{Re (t_{i})}^{Re(t_{j})}{\omega_{\bm p}(t)dt},$$
where $t_{i}$ and $t_{j}$ are the solutions of equation $\omega_{\bf p}(t)= \sqrt{m^{2}+p_{\perp}^{2}+(p_{x}-eA(t))^{2}}=0$, which denote the different turning points. $\theta_{(i,j)}$ represents the phase accumulated between different pairs of turning points, which is also called interference term. It is noticed that the turning points closest to the real $t$ axis dominate the number of created particles, while the distances of turning points along the real-axis direction dominate the interference effect in the momentum spectrum. The detailed explanation of the turning point structure to the momentum spectrum and the particle number will be analyzed and discussed in Sec. \ref{result3}.

\subsection{Model for the external field}\label{field}

We investigate pair production in $1+1$ dimensional spatially inhomogeneous electric field with the sinusoidal phase modulation, where the field model can be described as \cite{Gong:2019sbw}
\begin{equation}\label{FieldMode}
\begin{aligned}
E\left(x,t\right)
&=E_{0} f \left( x \right ) g\left( t \right )\\
&=E_{0} \exp \left(-\frac{x^{2}}{2 \lambda^{2}} \right ) \exp \left(-\frac{t^{2}}{2 \tau^{2}} \right ) \cos(\omega t + b \sin(\omega_{m} t ) ),
\end{aligned}
\end{equation}
where $E_{0}$ denotes the field strength, $\lambda$ is the spatial scale, $\tau$ is the pulse duration, $\omega$ represents the central frequency, $b$ and $\omega_{m}$ are the amplitude and frequency of the phase modulation, respectively. The nonzero modulated frequency $\omega_{m}$ and modulated amplitude $b$ lead to the time-dependent effective frequency $\omega_{\text{eff}}(t)=\omega+b \omega_{m} \cos(\omega_{m} t )$. In order to keep the modulation within a reasonable range, we set $|b \omega_{m} \cos(\omega_{m} t )|\leq\alpha\omega$ with $0<\alpha<1$. Because of $|b \omega_{m} \cos(\omega_{m} t )|_{max}=b \omega_{m}$, the inequality $b \omega_{m}\leq\alpha\omega$ can be derived. We can further obtain the relationship $b\leq\alpha\omega/\omega_{m}$, which also indicates that the upper and lower limits of the modulated amplitude can be obtained. Without losing the generality, we select the regime of $0\leq\alpha\leq0.9$ and the modulated frequency $\omega_{m}\approx(1/5\sim1/10)\omega$, meanwhile, it is noted that when we choose the maximum modulated amplitude, the minimum value of modulated frequency is considered, i.e., $\omega_{m}=1/10\omega$. For high frequency field, the field parameters are set to $E_{0}=0.3E_{\mathrm{cr}}$, $\omega=0.5$, $\tau=100$, therefore the corresponding maximum values of modulated amplitude and frequency can be selected as $b=0.9\omega/\omega_{m}=9$ and $\omega_{m}=1/5\omega=0.1$ , respectively. For low frequency field, we set $E_{0}=0.5E_{\mathrm{cr}}$, $\omega=0.1$, $\tau=25$, and the corresponding maximum values of modulated amplitude and frequency are chosen as $b=0.9\omega/\omega_{m}=9$ and $\omega_{m}=1/5\omega=0.02$, respectively.

The model in Eq.~\eqref{FieldMode} may be viewed as a field generated in the antinode of the standing-wave mode. Since the particles are mainly produced in the $x$-axis direction of the electric field, $p_{\perp}=0$ can be assumed.

\section{High frequency field}\label{results}

In this section, we investigate the effects of sinusoidal phase modulation on the momentum spectrum and the reduced particle number of the created particles in high frequency inhomogeneous field. In this case, the field parameters are set to $E_{0}=0.3E_{\mathrm{cr}}$, $\omega=0.5$, $\tau=100$, which corresponds to the multiphoton-dominant pair production process.

\subsection{Momentum spectrum}

We first study the influence of the modulated amplitude and frequency on the momentum spectrum for various spatial scales, respectively.

\subsubsection{Modulated amplitude}

When the modulated frequency is fixed $\omega_{m}=0.1\omega=0.05$, the momentum spectrum for different spatial scales with various modulated amplitude $b$ is shown in Fig.~\ref{1}. At the large spatial scale $\lambda=1000$, when $b=0$, there is a weak oscillation on the momentum spectrum, but we observed the spectrum symmetry, as shown in Fig.~\ref{1}(a). Since in the quasihomogeneous limit, i.e., $\lambda=1000$, the electric field model proposed in Eq.~\eqref{FieldMode} can be written as an even function of only time dependent as $E(x,t)\approx E(t)=E(-t)$, which leads to the symmetry of the momentum spectrum. Actually this symmetry is due to the fact that the DHW equations for getting the momentum distribution function are invariant under time reversal. Under time reversal, the time $t$ and the momentum $p_{x}$ change sign, the $x$ does not change sign. Because $\Omega(-p_{x})=\sqrt{(-p_{x})^{2}+m^2}=\Omega(p_{x})$, we can know from Eq.~\eqref{vacuum-initial} that ${\mathbbm{s}}^{v}(x,p_{x},t) = {\mathbbm{s}}^{v}(x,-p_{x},-t)$ and ${\mathbbm{v}_{1}}^{v}(x,p_{x},t) = - {\mathbbm{v}_{1}}^{v}(x,-p_{x},-t)$. At the same time, the odd/even property of other physical quantities that affect the momentum distribution function can be obtained by Eq.~\eqref{pseudoDiff} and Eqs.~(7)-(10), as shown in Table~\ref{Table 1}. Finally, it is found that the form of the DHW equations stays invariant under time reversal, which ensures the symmetry of the momentum spectrum.

\begin{table}[H]
\caption{The odd/even property, i.e., $-/+$ sign presentation, of the physical quantity under time reversal.}
\centering
\begin{ruledtabular}
\begin{tabular}{cccc}
physical quantity  &$(t,p_{x},x,\Omega,E)$  &$(\partial_{t},\partial_{p_{x}},\partial _{x},D_t)$  &$(\mathbbm{s},\mathbbm{v}_{1},\mathbbm{p},\mathbbm{v}_{0})$\\
\hline
$-/+$  &$(-,-,+,+,+)$     &$(-,-,+,-)$      &$(+,-,+,+)$ \\
\end{tabular}
\end{ruledtabular}
\label{Table 1}
\end{table}

For small modulated amplitude, the momentum spectrum is approximately symmetrical and presents obvious oscillation, as shown in Fig.~\ref{1}(b). Since the effective frequency of external field {$\omega_{\text{eff}}(t)=\omega+b \omega_{m} \cos(\omega_{m} t )$ is increased significantly as modulated amplitude increases, resulting in the enhancement of oscillation. It implies that there are remarkable interference effect on the momentum spectrum. For large modulated amplitude, one can see that the obvious oscillation and the mergence of two dominant peaks on the momentum spectrum, as shown in Figs.~\ref{1}(c) and (d). Meanwhile, the symmetry of momentum spectra is destroyed severely. Since when modulated amplitude is large, electric field presents the highly amplitude oscillation so that the degree of spatial quasihomogeneous is reduced, i.e., the validity of $E(x,t)\approx E(t)$ is lost more and more, which results in the symmetry destroy of the momentum spectrum.

When the spatial scale decreases to $\lambda=10$, there is no pronounced oscillation in the momentum spectrum for $b=0$, but it is approximately symmetrical, while the symmetry is destroyed for small modulated amplitude, as shown in Figs.~\ref{1}(a) and (b). Because the finite spatial scale of electric field prevents particle production from being dominated by the temporal pulse structure, which leads to the asymmetry of the momentum spectrum. For large modulated amplitude, we can observe that the momentum spectra show strong oscillation and obvious asymmetry in Figs.~\ref{1}(c) and (d), meanwhile, the momentum distribution range is broadened. Since the finite laser pulse seems to prevent the coherent superposition of the particle trajectory, and the corresponding interference pattern is disturbed, which further leads to a broadening of the momentum distribution.

\begin{figure}[H]%\suppressfloats
\begin{center}
%max\includegraphics[width=\textwidth]{Fig2}
\includegraphics[width=\textwidth]{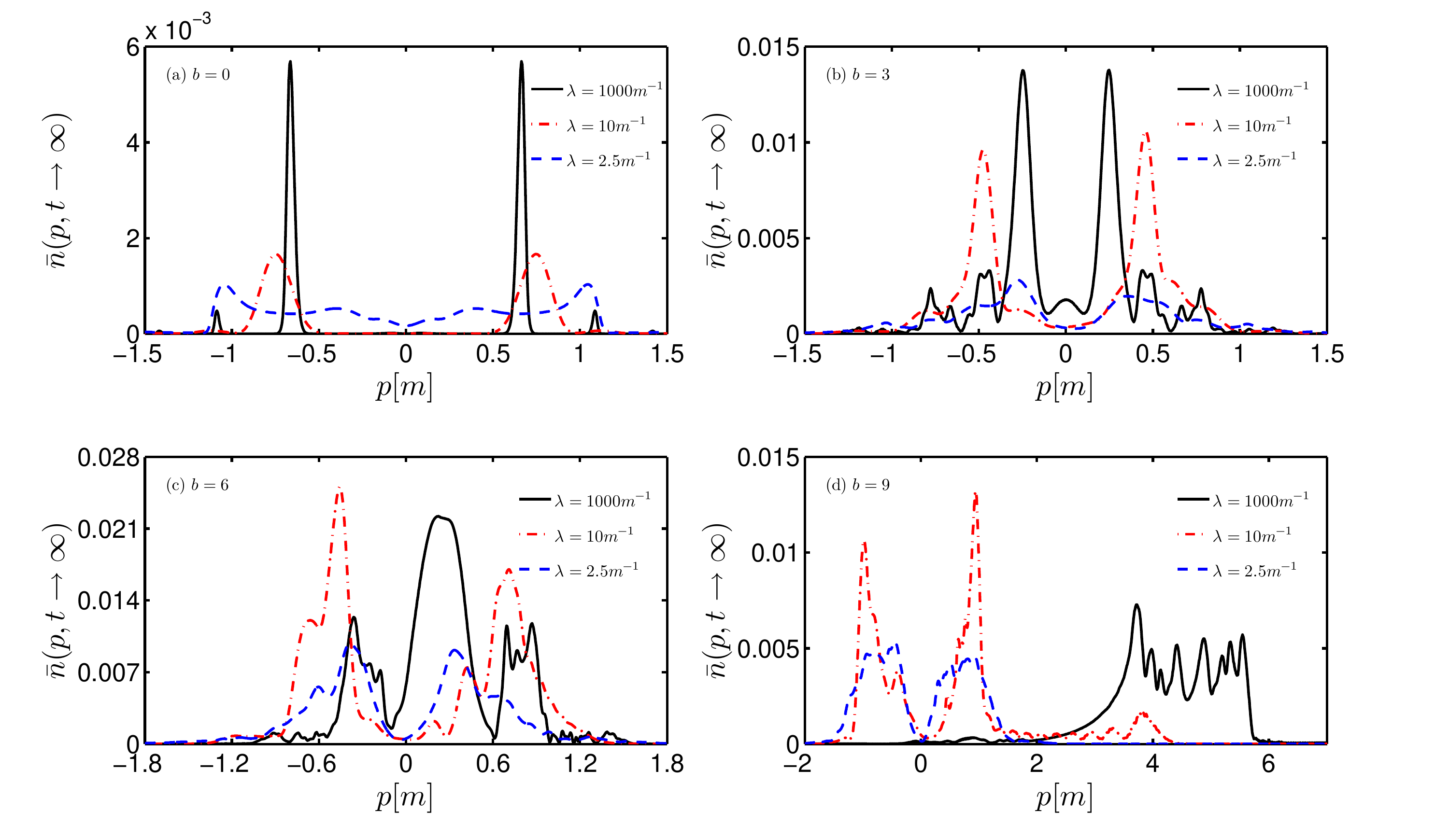}
\end{center}
\setlength{\abovecaptionskip}{-0.3cm}
\setlength{\belowcaptionskip}{-0.3cm}
\caption{(color online). Reduced momentum spectrum for various modulated amplitude values in high frequency field with different spatial scales when modulated frequency is $\omega_{m}=0.05$. The modulated amplitude values are $b=0, 3, 6$ and $9$, respectively. Other field parameters are $E_{0}=0.3E_{\mathrm{cr}}$, $\omega=0.5$, $\tau=100$.}
\label{1}
\end{figure}

At the extremely small spatial scale $\lambda=2.5$, when $b=0$, the weak oscillation occurs on the momentum spectrum, meanwhile, an approximate symmetry can be observed, as shown in Fig.~\ref{1}(a). The influence of electric field focusing on the small spatial scale is so small that the Eq.~\eqref{pseudoDiff} can be written as $D_t\approx\partial_{t}$. Meanwhile, it is found that the odd/even property of physical quantities that affect the momentum spectrum under time reversal are the same as those in Table~\ref{Table 1}. Therefore, the form of the DHW equations still stays invariant, which leads to an approximate symmetry of the momentum spectrum. With modulated amplitude increases, one can see that the symmetry is destroyed gradually because the validity of $D_t\approx\partial_{t}$ is lost more and more. On the other hand, however, the obvious oscillation appears in the momentum spectrum, as shown in Figs.~\ref{1}(b), (c) and (d).

\subsubsection{Modulated frequency}

When the modulated amplitude is fixed $b=0.1\omega/\omega_{m}=1$, the momentum spectrum for different spatial scales with various modulated frequency $\omega_{m}$ is displayed in Fig.~\ref{2}. It is noted that the result of $\omega_{m}=0$ is the same as that in Fig.~\ref{1}(a) with $b=0$, which indicates the electric field without modulation. At the large spatial scale $\lambda=1000$, the weak oscillation can be observed even for small modulated frequency, as shown in Figs.~\ref{2}(a) and (b), which may be viewed as the interference effect between the created particles. Compared with the case of $\omega_{m}=0$, the maximum peak values of the momentum spectra are increased about $5$ times. Since we take the Fourier transform of $E\left(t\right)=E_{0} \exp \left(-\frac{t^{2}}{2 \tau^{2}} \right ) \cos(\omega t + b \sin(\omega_{m} t ) )$ and find that for $\omega_{m}=0$, only one dominant frequency occurs on the Fourier spectrum, while for $\omega_{m}=0.05$ and $0.07$, there are not only one primary frequency but also three pairs of symmetrical subfrequencies, which provides a great contribution to the external field frequency and leads to more energy to produce more particles. Particularly, some frequency spectrum structures can play a dynamically assisted role, which can enhance significantly the number of created particles. For large modulated frequency, one can see that strong oscillation appears, while compared with the case of small modulated frequency, there are no significant increase of the maximum peak values, as shown in Figs.~\ref{2}(c) and (d). On the other hand, compared with the Fig.~\ref{1}, it is found that there is a good symmetry on the momentum spectrum with modulated frequency, while it is destroyed severely when the modulated amplitude becomes large. It indicates that the momentum spectrum is more sensitive to modulated amplitude.

When the spatial scale reduces to $\lambda=10$ and $\lambda=2.5$, for small modulated frequency, there is no oscillation but one can see that the momentum spectrum presents an approximate symmetry, see Figs.~\ref{2}(a) and (b). For large modulated frequency, we observed weak oscillation but the symmetry of momentum spectrum is destroyed, as shown in Figs.~\ref{2}(c) and (d). Moreover, there are some different phenomena on the momentum spectrum at $\lambda=10$ and $\lambda=2.5$. For $\lambda=10$, compared to the case of $\lambda=1000$, the dominant peaks of all the momentum spectra in Fig.~\ref{2} are shifted to the direction of large momentum, which can be explained by ponderomotive force \cite{Kohlfurst:2017hbd}. Since the ponderomotive force is inversely proportional to the size of spatial scale, i.e., the smaller the spatial scale, the stronger the ponderomotive force is. Therefore, the dominant peaks of the momentum spectra at spatial scale $\lambda=10$ are further pushed away from the center compared with the case of $\lambda=1000$. For $\lambda=2.5$, compared to the case of $\lambda=10$, the momentum peaks are not pushed away from the center in Fig.~\ref{2}. Since the highly inhomogeneous oscillation caused by the increasing of modulated frequency decreases the corresponding effect of ponderomotive force, which leads to that the particles are not pushed towards the regions of low field strength.

\begin{figure}[H]%\suppressfloats
\begin{center}
\includegraphics[width=\textwidth]{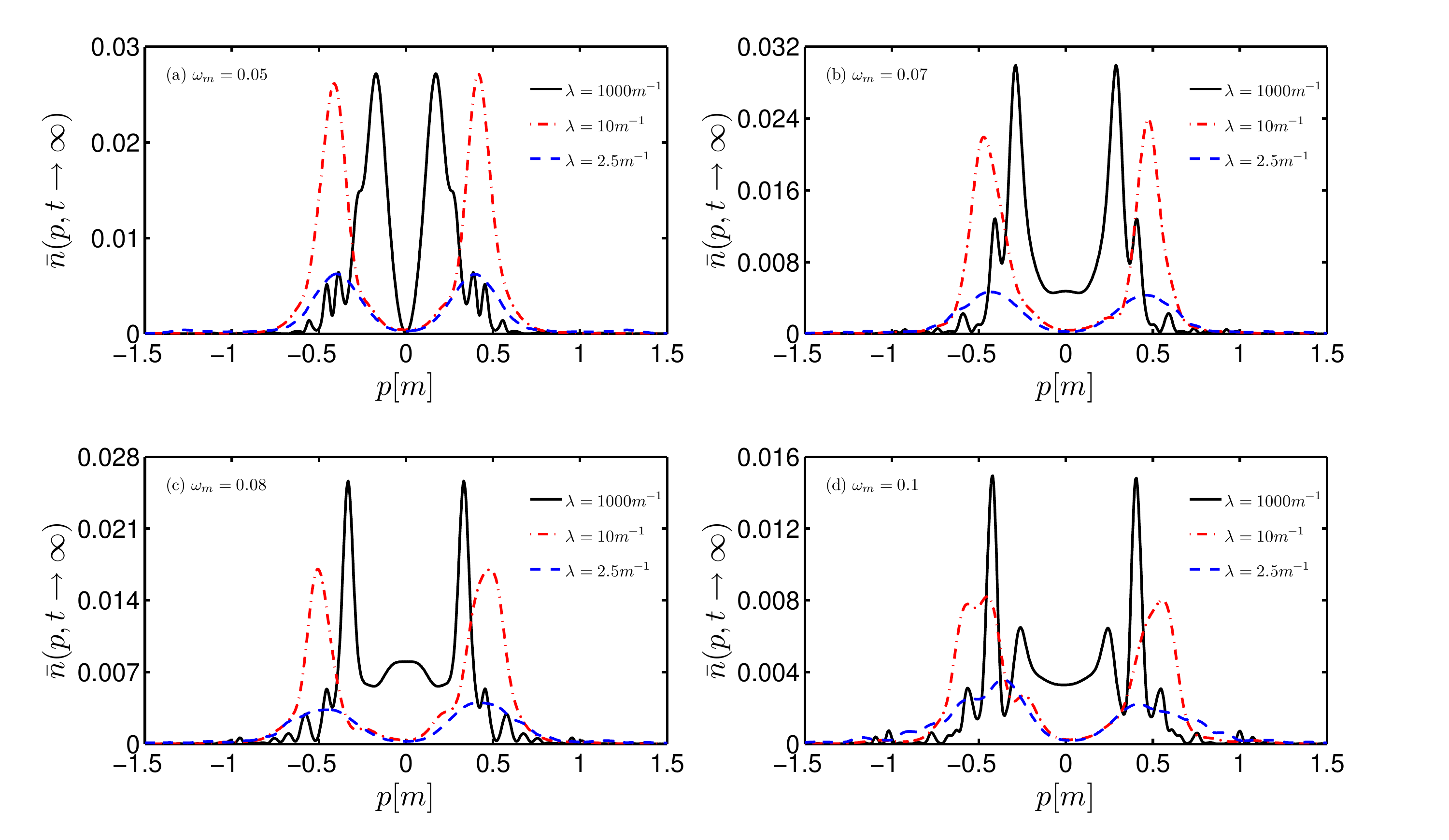}
\end{center}
\setlength{\abovecaptionskip}{-0.3cm}
\setlength{\belowcaptionskip}{-0.3cm}
\caption{(color online). Reduced momentum spectrum for various modulated frequency values in high frequency field with different spatial scales when modulated amplitude is $b=1$. The modulated frequency values are $\omega_{m}=0.05, 0.07, 0.08$ and $0.1$, respectively. Other field parameters are the same as in Fig.~\ref{1}.}
\label{2}
\end{figure}

\subsection{Reduced particle number}\label{resultB}

In this subsection, we study the effect of modulation parameters on the reduced particle number for various spatial scales, in the different cases, such as modulating only in amplitude, modulating only in frequency and modulating in both amplitude and frequency.

Figures~\ref{3}(a) and (b) show the reduced particle number dependence on spatial scales for various modulated amplitude and frequency, respectively. It can be seen that when the modulation parameters (either the modulated amplitude or frequency) are fixed, with increasing spatial scale, the reduced particle number is increased rapidly at small spatial scales, while it tends to be a constant at large spatial scales. Because at small spatial scales, the electric field strength becomes large with spatial scale, more particles are created in the whole range of the electric field. It is noted that for certain modulated frequency $\omega_{m}=0.07$ and $\omega_{m}=0.1$, the particle number is enhanced in the range of $1.6<\lambda<7$, while it is decreased in the range of $7<\lambda<9$, as shown in Fig.~\ref{3}(b). It means that the reduced particle number is more sensitive to modulated frequency, this reason will be discussed separately in the following section.

\begin{figure}[H]%\suppressfloats
\begin{center}
\includegraphics[width=\textwidth]{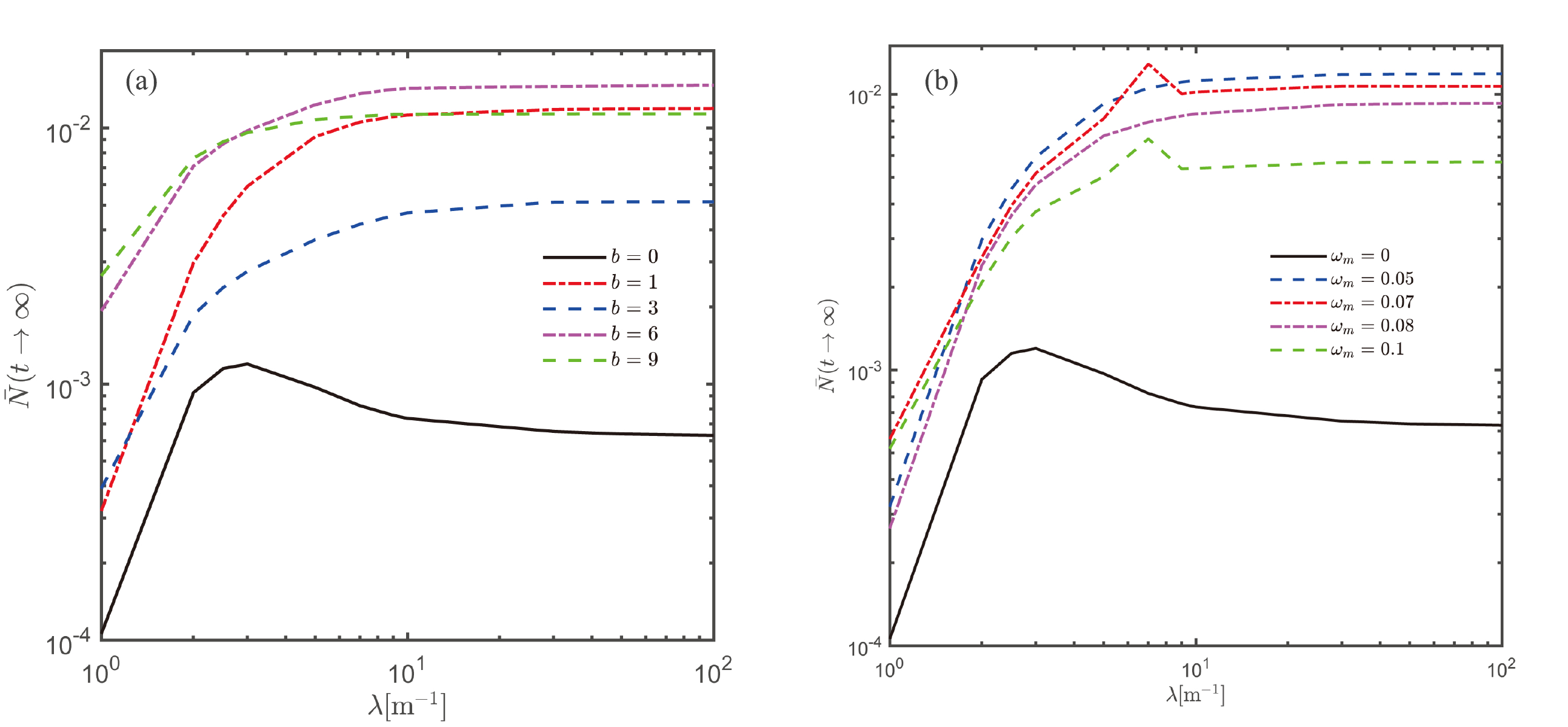}
\end{center}
\setlength{\abovecaptionskip}{-0.3cm}
\setlength{\belowcaptionskip}{-0.3cm}
\caption{(color online). Reduced particle number dependence on spatial scales for different modulated frequency and amplitude parameters in high frequency field. Panel (a): The plot is for the change of modulated amplitude with $\omega_{m}=0.05$. Panel (b): The plot is for the change of modulated frequency with $b=1$. Other field parameters are the same as in Fig.~\ref{1}.}
\label{3}
\end{figure}

When the spatial scale is fixed, the reduced particle number is enhanced significantly for various modulated amplitude and frequency. At large spatial scales, compared to the case of electric field without modulation, the particle number is increased rapidly at about one order of magnitude by either the large modulated amplitude or small modulated frequency, as shown in Fig.~\ref{3}. At small spatial scales, the enhancement of the particle number for various modulated amplitude and frequency is different. With modulated amplitude, the particle number can be increased at about one order of magnitude, as shown in Fig.~\ref{3}(a), while for modulated frequency, it is enhanced about $5$ times, as shown in Fig.~\ref{3}(b). Therefore, it indicates that the change of modulated amplitude is beneficial for the pair production. Meanwhile, when spatial scale is small, for small modulated amplitude, the particle number is enhanced rapidly with spatial scale, while there is no significant increase for large modulated amplitude, as shown in Fig.~\ref{3}(a). Since when modulated amplitude is large, the particle production process is dominated by multiphoton absorption that is less influenced by the spatial scale.

\begin{figure}[H]%\suppressfloats
\begin{center}
%max\includegraphics[width=\textwidth]{Fig2}
\includegraphics[scale=0.55]{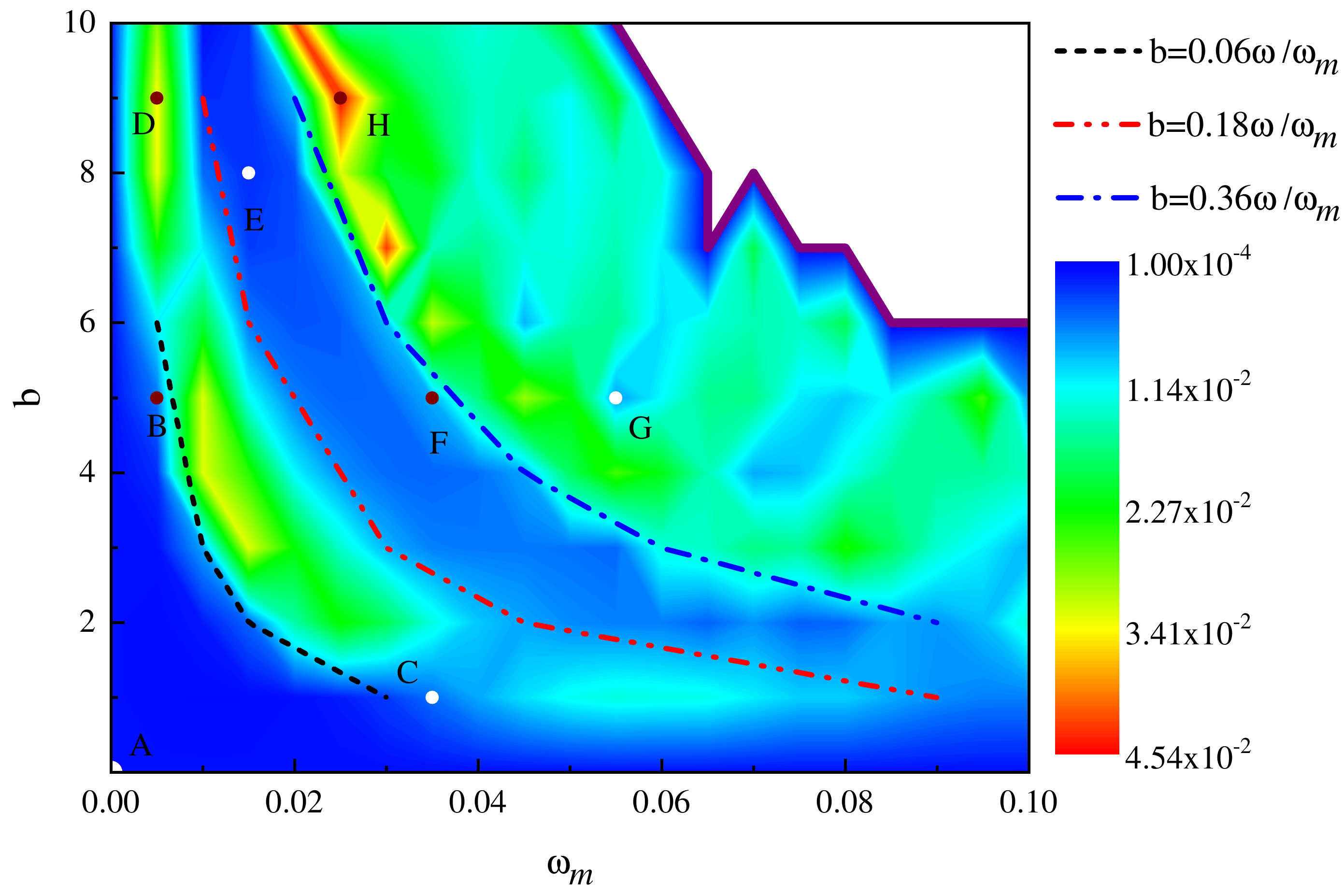}
\end{center}
\setlength{\abovecaptionskip}{-0.3cm}
\setlength{\belowcaptionskip}{-0.3cm}
\caption{(color online). Contour plots of the reduced particle number versus the modulated frequency and amplitude at spatial scale $\lambda=100$. Other field parameters are $E_{0}=0.3E_{\mathrm{cr}}$, $\omega=0.5$, $\tau=100$. Note that the blank area separated by the purple solid line is beyond the modulation range of $\alpha=b\omega_m/\omega<1$.}
\label{4}
\end{figure}

In order to study the effect of modulated frequency and amplitude on the reduced particle number more comprehensively, we present the contour plot as shown in Fig.~\ref{4}. Due to limited computational resources, we select an intermediate spatial scale $\lambda=100$ to study. One can see that there are several different regions on the diagram, which are divided by three typical curves, where these curves represent $b=0.06\omega/\omega_{m}$, $b=0.18\omega/\omega_{m}$ and $b=0.36\omega/\omega_{m}$, respectively. It is found that in the modulation range of $b\leq0.06\omega/\omega_{m}$, the reduced particle number is not sensitive to the modulation parameters, while it presents an obvious variation in the modulation range of $0.06\omega/\omega_{m}<b\leq0.18\omega/\omega_{m}$. Since when $b\leq0.06\omega/\omega_{m}$, the maximum value of effective frequency is $\omega+0.06\omega$, which has less distinction to the original center frequency. We also observe that when the modulation range is $0.18\omega/\omega_{m}<b<0.36\omega/\omega_{m}$, there is no pronounced change for the particle number, while it presents a significant variation in the range of $b\geq0.36\omega/\omega_{m}$. Moreover, the significant maximum (wine dot) and minimum (white dot) values of the reduced particle number on the different regions can be obtained, as shown in Table~\ref{Table 3}. It is found that when the modulation range are $0.06\omega/\omega_{m}<b\leq0.18\omega/\omega_{m}$ and $b\geq0.36\omega/\omega_{m}$, the reduced particle number is enhanced significantly. Particularly, in the range of $b\geq0.36\omega/\omega_{m}$, the maximum value of particle number is about $70$ times larger than that without modulation. Interestingly, we find that the same reduced particle number can be got through different set of modulation parameters.

\begin{table}[H]
\caption{The maximum and minimum values of the reduced particle number in different modulation ranges marked in Fig.~\ref{4}.}
\centering
\begin{ruledtabular}
\begin{tabular}{ccc}
modulation range   &modulation parameters $(\omega_{m},b)$  &reduced particle number\\
\hline
$b\leq0.06\omega/\omega_{m}$     &$A(0,1)$      &$6.30\times10^{-4}$ \\
                              &$B(0.005,5)$      &$5.19\times10^{-3}$\\
\hline
$0.06\omega/\omega_{m}<b\leq0.18\omega/\omega_{m}$ &$C(0.035,1)$     &$4.88\times10^{-3}$ \\
                                                   &$D(0.005,9)$      &$3.42\times10^{-2}$ \\
\hline
$0.18\omega/\omega_{m}<b<0.36\omega/\omega_{m}$ &$E(0.015,8)$      &$1.98\times10^{-3}$ \\
                                                 &$F(0.035,5)$   &$7.85\times10^{-3}$ \\
\hline
$b\geq0.36\omega/\omega_{m}$                        &$G(0.055,5)$   &$7.07\times10^{-3}$ \\
                                                 &$H(0.025,9)$  &$4.54\times10^{-2}$ \\
\end{tabular}
\end{ruledtabular}
\label{Table 3}
\end{table}

To study whether there are optimal modulation parameters for $e^{-}e^{+}$ pair production in high frequency field with sinusoidal phase modulation, we further explore the variation of the reduced particle number in the following two cases at the spatial scale $\lambda=100$, as shown in Fig.~\ref{5}(a). One case is that the electric field does not have any modulation, only the central frequency $\omega$. The other case is that the electric field has both central frequency and modulation, where the central frequency is $\omega=0.5$, and the modulation is that when the modulated frequency is fixed $\omega_{m}=0.01$, but the modulated amplitude $b$ changes. It is found that the trends of the results in the above two cases are almost identical. Since in the second case, we perform the Fourier transform of the time-dependent electric field and regard the frequency with the largest amplitude on the frequency spectrum as the original center frequency of the electric field without modulation, which makes the trends almost identical. Moreover, in the first case, it is found that the reduced particle number is extremely sensitive to the central frequency of the external field and presents an obvious nonlinear variation. In particular, when the central frequency is $\omega=0.69$, the particle number reaches the maximum value, i.e., $\bar{N}=0.1331$, and it is enhanced significantly by more than two orders of magnitude compared to the case without modulation, i.e., $\omega=0.5$. In the second case, we can also observe the sensitivity of the reduced particle number to the modulated amplitude and find that when modulated frequency $\omega_{m}=0.01$ and modulated amplitude $b=20.7$, the reduced particle number reaches the maximum value, i,e., $\bar{N}=0.1367$, which is almost the same as the maximum value in the first case.

\begin{figure}[H]
\centering
\subfigure{\includegraphics[width=7.5cm]{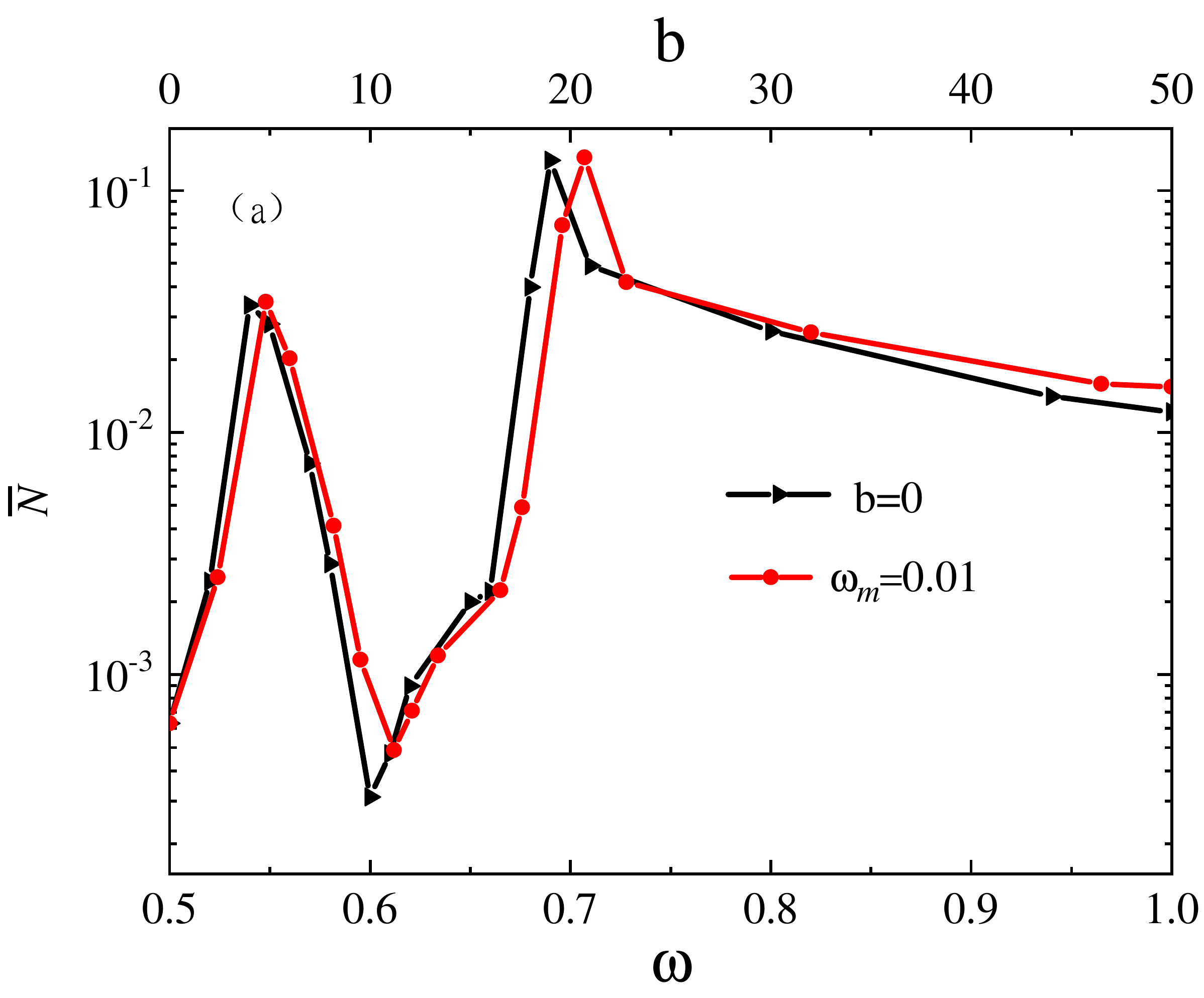}}
\hspace{7mm}
\subfigure{\includegraphics[width=7.52cm]{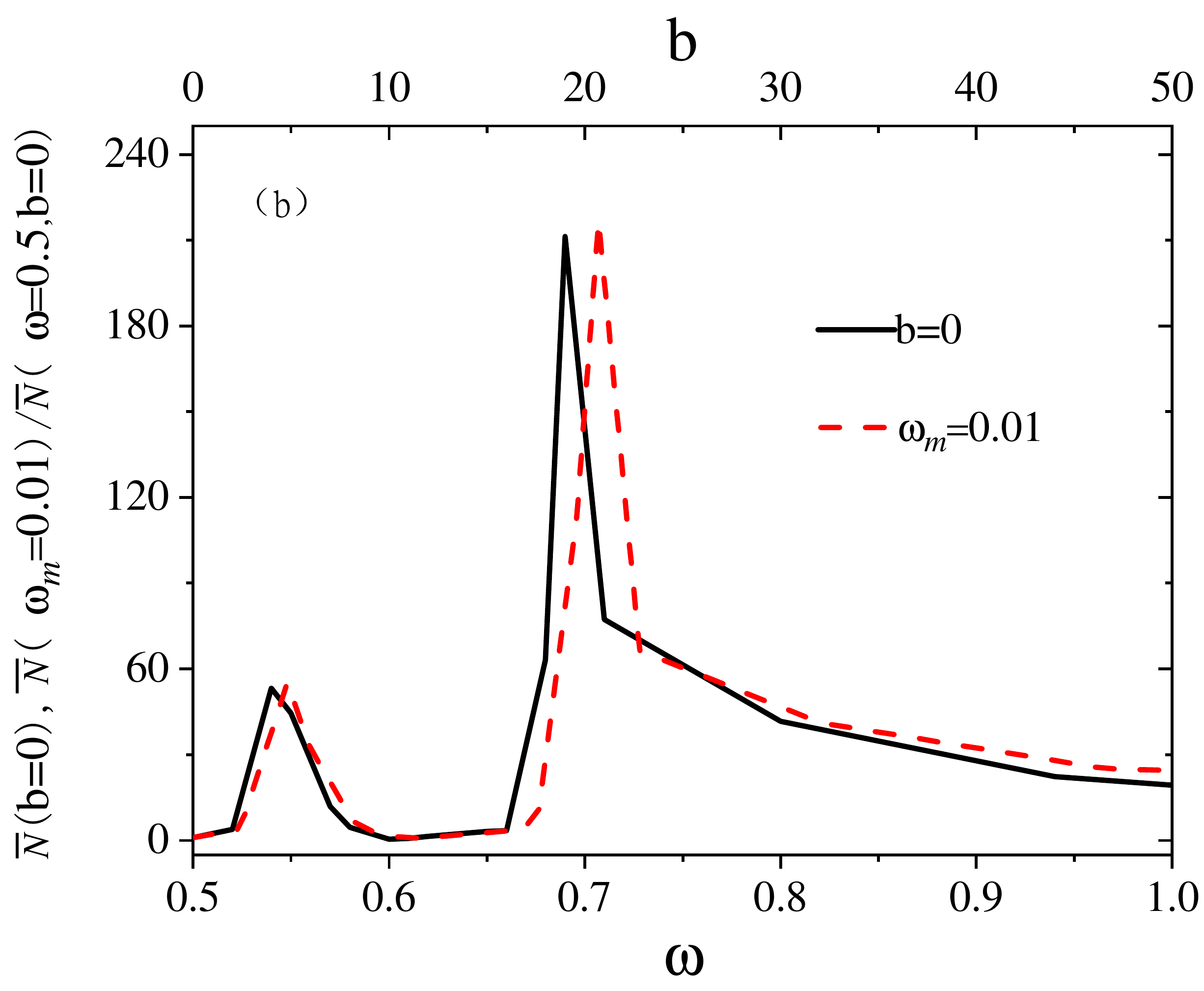}}
\setlength{\abovecaptionskip}{-0.3cm}
\setlength{\belowcaptionskip}{-0.3cm}
\caption{(color online). (a) Reduced particle number as a function of the central frequency of the external field (black  line) when $b=0$ and the modulated amplitude (red line) when $\omega_m=0.01$ is fixed, respectively. Other field parameters are the same as in Fig.~\ref{1}. (b) The ratio of $\bar{N}(b=0)/\bar{N}(\omega=0.5,b=0)$,  $\bar{N}(\omega_{\text{m}}=0.01)/\bar{N}(\omega=0.5,b=0)$ as a function of the central frequency (black line) and the modulated amplitude (red line), respectively.}
\label{5}
\end{figure}

Figure \ref{5}(b) shows the change in the ratio of the reduced particle number in the above two cases and the particle number under the electric field without modulation. In the first case, one can see that when the central frequency of the external field is $\omega=0.69$, the ratio reaches the maximum value, i.e., $\bar{N}(b=0)/\bar{N}(\omega=0.5,b=0)\approx211$, while in the second case, when modulated frequency $\omega_{m}=0.01$ and modulated amplitude $b=20.7$, the maximum value of the ratio is $\bar{N}(\omega_{\text{m}}=0.01)/\bar{N}(\omega=0.5,b=0)\approx217$. It indicates that in the two case, the reduced particle numbers are enhanced significantly about $200$ times compared to the case without modulation, i.e., $\omega=0.5$. Therefore, we can obtain that $\omega=0.69$ is the optimal value of the central frequency of the external field to get the largest reduced particle number, meanwhile, $\omega_{m}=0.01$ and $b=20.7$ are the optimal values of the modulated frequency and the modulated amplitude to get the largest reduced particle number.

\section{LOW Frequency Field}\label{result2}

In this section, the influence of sinusoidal phase modulation on the momentum spectrum and the reduced particle number of the created particles in low frequency inhomogeneous field is studied. In this case, we set $E_{0}=0.5E_{\mathrm{cr}}$, $\omega=0.1$, $\tau=25$, which corresponds to the tunneling-dominated pair production process.

\subsection{Momentum spectrum}

We study the effect of the modulated amplitude and frequency on the momentum spectrum for various spatial scales, respectively.

\subsubsection{Modulated amplitude}

When the modulated frequency is fixed $\omega_{m}=1/10\omega=0.01$, the momentum spectrum for different spatial scales with various modulated amplitude $b$ is shown in Fig.~\ref{6}. At the large spatial scale $\lambda=500$, when $b=0$, one can see that the bell-shaped momentum spectrum presents a weak oscillatory structure, as shown in Fig.~\ref{7}(a). The result is consistent with the case of $b=0$ in Fig.~\ref{6}(a) of Ref. \cite{Ababekri:2020}. With modulated amplitude, there are obvious oscillation on the momentum spectrum, meanwhile, the maximum peak values of the momentum spectra are increased significantly, as shown in Figs.~\ref{6}(b), (c) and (d). Since in the quasihomogeneous limit ($\lambda=500$), we know from the Fourier transform of $E\left(t\right)=E_{0} \exp \left(-\frac{t^{2}}{2 \tau^{2}} \right ) \cos(\omega t + b \sin(\omega_{m} t ) )$ that the dominant frequency on the frequency spectrum is enhanced with modulated amplitude, which leads that the corresponding Keldysh adiabaticity parameter $\gamma=m\omega/eE$ becomes larger. It indicates that the pair production process tends to the multiphoton absorbtion region, resulting in a significant enhancement of the maximum peak value of the momentum spectrum.

\begin{figure}[h]%\suppressfloats
\begin{center}
\includegraphics[width=\textwidth]{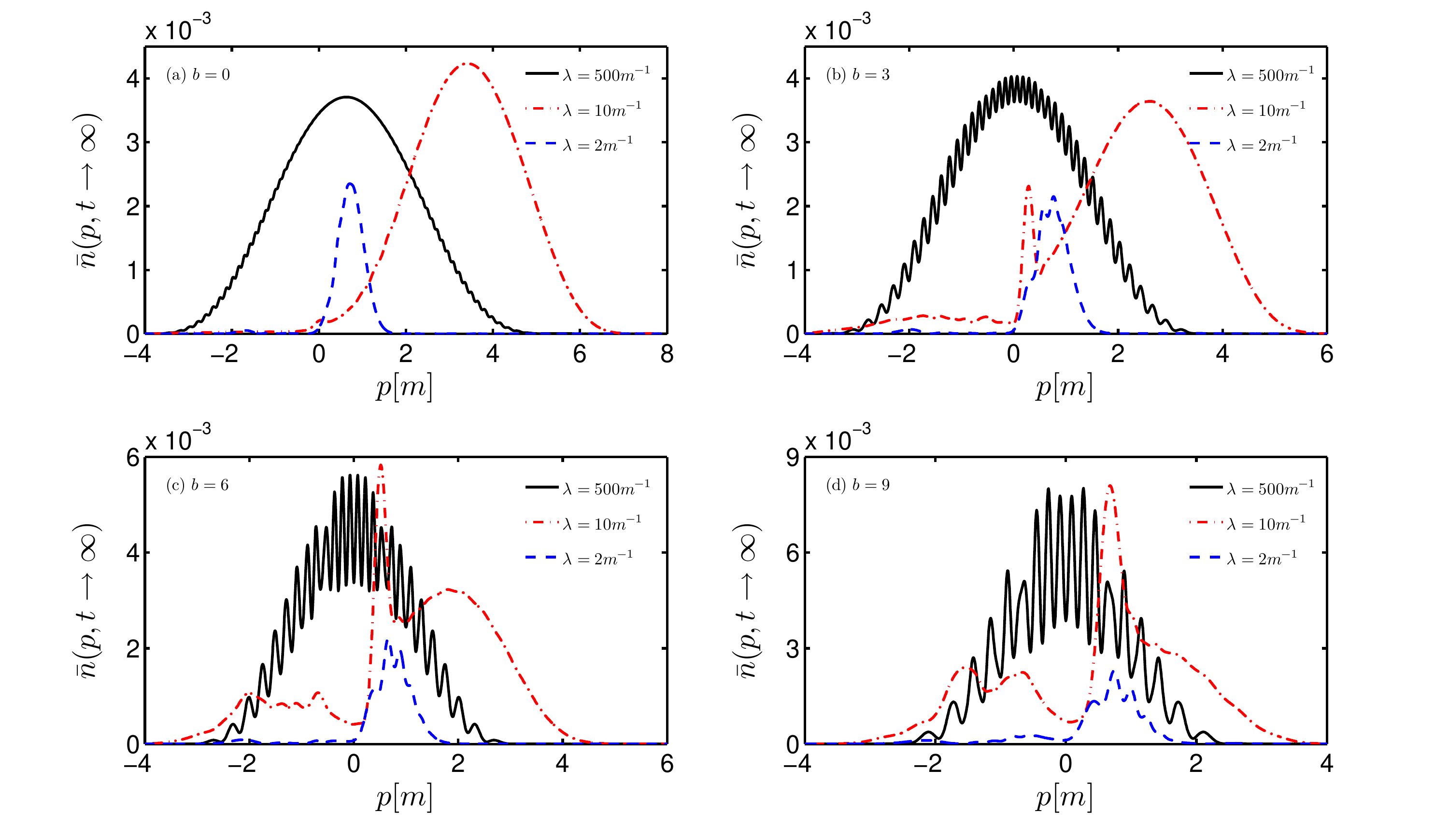}
\end{center}
\setlength{\abovecaptionskip}{-0.3cm}
\caption{(color online). Reduced momentum spectra for various  modulated amplitude values in low frequency field with different spatial scales when modulated frequency $\omega_{\text{m}}=0.01$. The modulated amplitude values are $b=0, 3, 6$ and $9$, respectively. Other field parameters are $E_0=0.5E_{\mathrm{cr}}$, $\omega=0.1$, $\tau=25$.}
\label{6}
\end{figure}

When the spatial scale is reduced to $\lambda=10$, there is no oscillation on the momentum spectrum for $b=0$, but it shifts to the larger momentum values, see Fig.~\ref{6}(a). Because the small spatial scale causes the electric field strength to decrease rapidly, so the created particles with large momentum can escape from the field region more easily. For small modulated amplitude, we observe weak oscillation on the momentum spectrum and find a distinct small peak near the vanishing momentum $p=0$, see Fig.~\ref{6}(b). For large modulated amplitude, there is an obvious oscillation on the momentum spectrum, meanwhile, it is found that the peak near the $p=0$ is enhanced significantly, which is larger than the momentum peak of $\lambda=500$, see Figs.~\ref{6}(c) and (d). It means that the strong nonlinear tunneling effect occurs on the pair creation. We also find that the symmetry of the momentum spectra in Fig.~\ref{6} is destroyed by the finite spatial scales of the electric field.

When the spatial scale is further decreased to $\lambda=2$, there is no obvious oscillation on the momentum spectrum for $b=0$, but it is approximately symmetrical, as shown in Fig.~\ref{6}(a). With modulated amplitude, the weak oscillation can be observed and the symmetry of momentum spectrum is destroyed, as shown in Figs.~\ref{6}(b), (c), and (d). Moreover, when $\lambda=2$, compared with the case of $\lambda=10$, the range of momentum distribution is reduced significantly in Fig.~\ref{6}. Since the work done by the external field is very small at extremely narrow spatial scale, which leads to a significant decrease of the particles created.

\subsubsection{Modulated frequency}

\begin{figure}[h]%\suppressfloats
\begin{center}
\includegraphics[width=\textwidth]{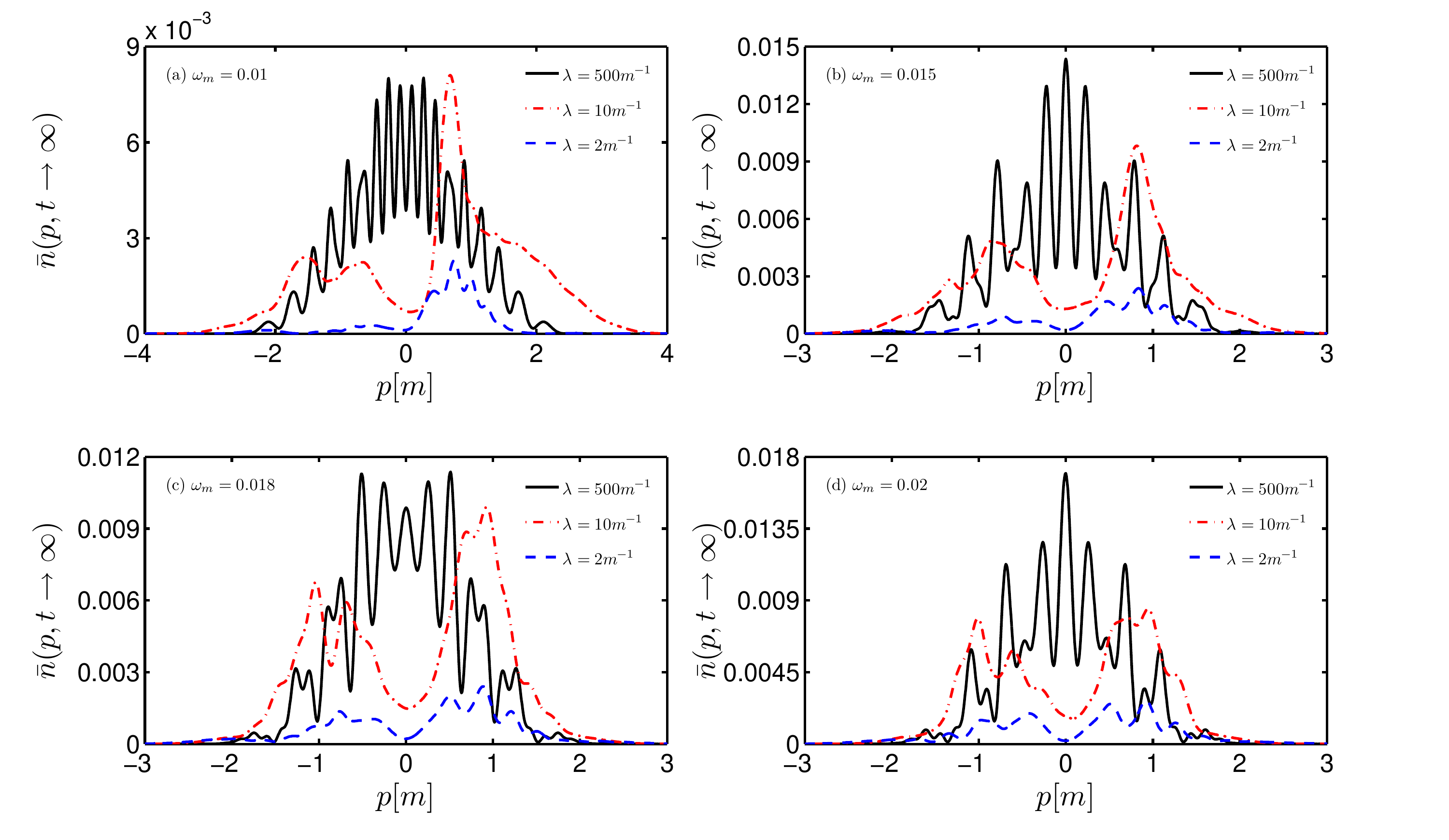}
\end{center}
\setlength{\abovecaptionskip}{-0.3cm}
\caption{(color online). Reduced momentum spectrum for various modulated frequency values in low frequency field with different spatial scales when modulated amplitude is $b=9$. The modulated frequency values are $\omega_{m}=0.01, 0.015, 0.018$ and $0.02$. Other field parameters are the same as in Fig.~\ref{6}.}
\label{7}
\end{figure}

When the modulated amplitude is fixed $b=0.9\omega/\omega_{\text{m}}=9$, the momentum spectrum for different spatial scales with various modulated frequency $\omega_{m}$ is shown in Fig.~\ref{7}. For small modulated frequency, one can see that there is an obvious oscillation on the momentum spectrum, see Fig.~\ref{7}(a). The complex oscillation may be understood as interference effect of pair creation by opposite signed large peaks and a series of small peaks in the low frequency field. With modulated frequency, the oscillation becomes more and more dispersive, but the maximum peak value of the momentum spectrum is enhanced greatly, as shown in Figs.~\ref{7}(b), (c) and (d). Interestingly, compared with the Fig.~\ref{6}, we find that there are different oscillatory structures on the momentum spectrum. When modulated frequency increases, the momentum spectrum shows dispersive oscillation, as shown in Fig.~\ref{7}, while with modulated amplitude, it presents intensive oscillation, as shown in Fig.~\ref{6}, which is related to the frequency spectrum structure. It is well known that the modulated amplitude dominates the amplitude of the frequency spectrum, while the modulated frequency determines the width of the spectrum.

When the spatial scale is reduced to $\lambda=10$ and $\lambda=2$, for small modulated frequency, one can see that the momentum spectrum presents obvious oscillation, but it is asymmetric, as shown in Fig.~\ref{7}(a). With modulated frequency, it shows an approximate symmetry and appears complex oscillation, as shown in Figs.~\ref{7}(b), (c) and (d), which is caused by the interference of produced particles from opposite field peaks. Moreover, compared with the case of $\omega_{\text{m}}=0$, the momentum distribution extends into the negative field region. Since the created particles with a certain momentum leave the external field region and miss the deceleration of the negative field peak.

\subsection{Reduced particle number}\label{resultD}

In this subsection, we investigate the effect of modulated frequency and amplitude on the reduced particle number for different spatial scales, as shown in Fig.~\ref{8}.

\begin{figure}[h]%\suppressfloats
\begin{center}
\includegraphics[width=\textwidth]{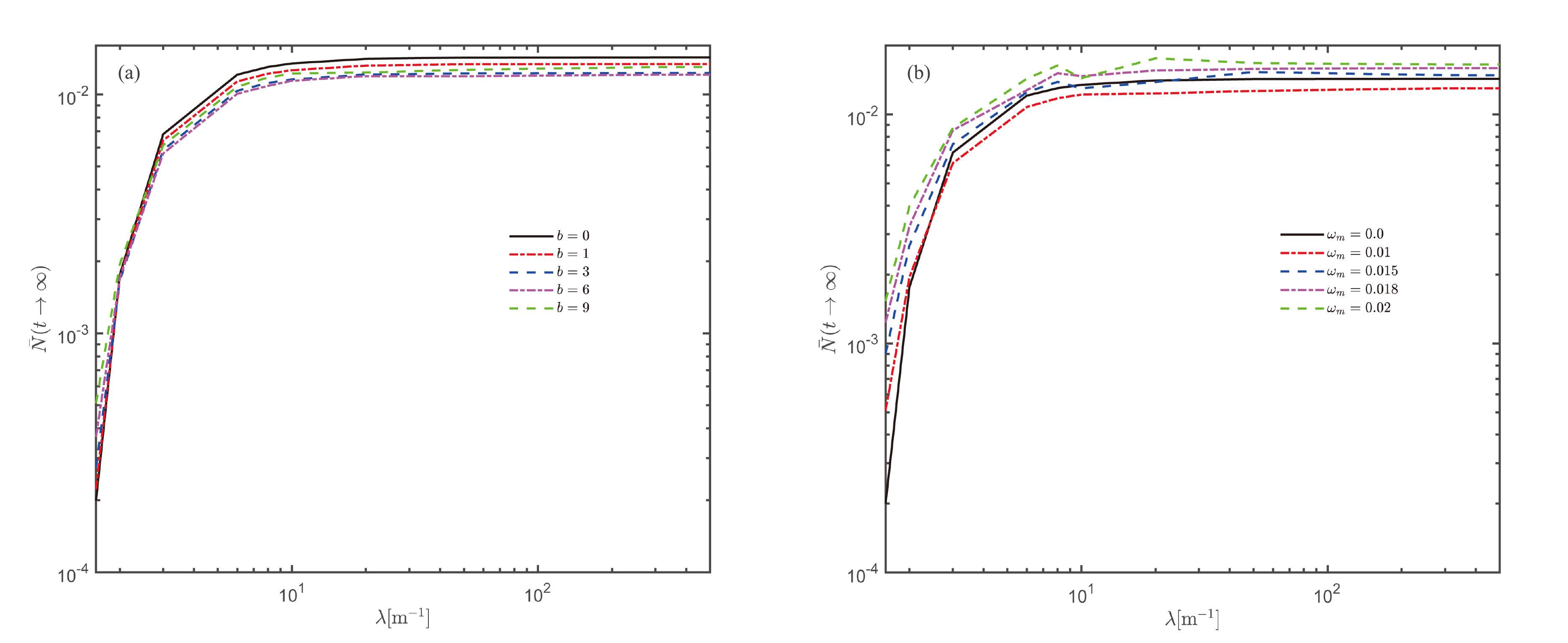}
\end{center}
\setlength{\abovecaptionskip}{-0.3cm}
\setlength{\belowcaptionskip}{-0.3cm}
\caption{(color online). Reduced particle number dependence on spatial scales for different modulated frequency and amplitude parameters in low frequency field. Panel (a): The plot is for the change of modulated amplitude when $\omega_{m}=0.01$. Panel (b): The plot is for the change of modulated frequency when $b=9$. Other field parameters are the same as in Fig. \ref{6}.}
\label{8}
\end{figure}

It can be seen from the Fig.~\ref{8} that when modulation parameters (either the modulated amplitude or frequency) are fixed, with spatial scale, the reduced particle number is enhanced significantly at small spatial scales, while it is almost unchangeable at large spatial scales. Interestingly, at small spatial scales, when the modulated frequency are $\omega_{m}=0.015$ and $\omega_{m}=0.02$, the reduced particle number shows a clear nonlinear variation, as shown in Fig.~\ref{8}(b). When spatial scale is fixed, both large modulated amplitude and large modulated frequency are beneficial to the increase of reduced particle number at small spatial scales. Compared to the case of electric field without modulation, in the case of modulated amplitude, the particle number is enhanced about $9$ times, see Fig.~\ref{8}(a), while in the case of modulated frequency, it is increased about $3$ times, see Fig.~\ref{8}(b). At large spatial scales, the reduced particle number is insensitive to modulated amplitude, while for the modulation frequency, it is enhanced by a small amount.

\section{DISCUSSION}\label{result3}

In this section, we employ the semiclassical WKB approach to qualitatively discuss the interference effect of the momentum spectrum. The nonlinear effect of the reduced particle number is discussed qualitatively by using the view point from the action of worldline instanton, meanwhile, it is analyzed briefly by the position distribution.

\subsection{Interference effect of the momentum spectrum}

According to the semiclassical WKB approach in the previous section, the interference effect of the momentum spectrum can be understood qualitatively by the location of turning points in the complex $t$ plane. Figure~\ref{9} shows the turning points distribution corresponding to the maximum peak values of the momentum spectrum for various modulated frequency. It can be seen from the Fig.~\ref{9}(a), when $\omega_{m}=0$, there are two pairs of turning points closest to the real $t$ axis, while for small modulated frequency, we observed five pairs of turning points are closest to the real $t$ axis, see Fig.~\ref{9}(b). Since the more pairs of turning point closest to the real $t$ axis means that the remarkable interference effect will appear in the momentum spectrum. Therefore, there is weak interference on the momentum spectrum in Fig.~\ref{1}(a), while it presents an obvious interference effect in Fig.~\ref{2}(a). With modulated frequency, one can see that the neighbouring distance of turning points is shorter and shorter, which depends on the modulated frequency, therefore, it leads to the more turning points near the real $t$ axis, as shown in Figs.~\ref{9}(c) and (d). Correspondingly, the interference effect of the momentum spectra becomes more and more remarkable, as shown in Figs.~\ref{2}(b) and (d). Compared with the case of Fig.~\ref{9}(a), the turning points in Figs.~\ref{9}(b), (c) and (d) present the closer distance to the real $t$ axis, which implies that the particles production rate in the modulating case is larger than that without modulation. Moreover, all the turning point structures in Fig.~\ref{9} are approximately symmetrical, which results in a symmetry of the corresponding momentum spectrum.

\begin{figure}[H]
\centering
\subfigure{\includegraphics[width=6.8cm]{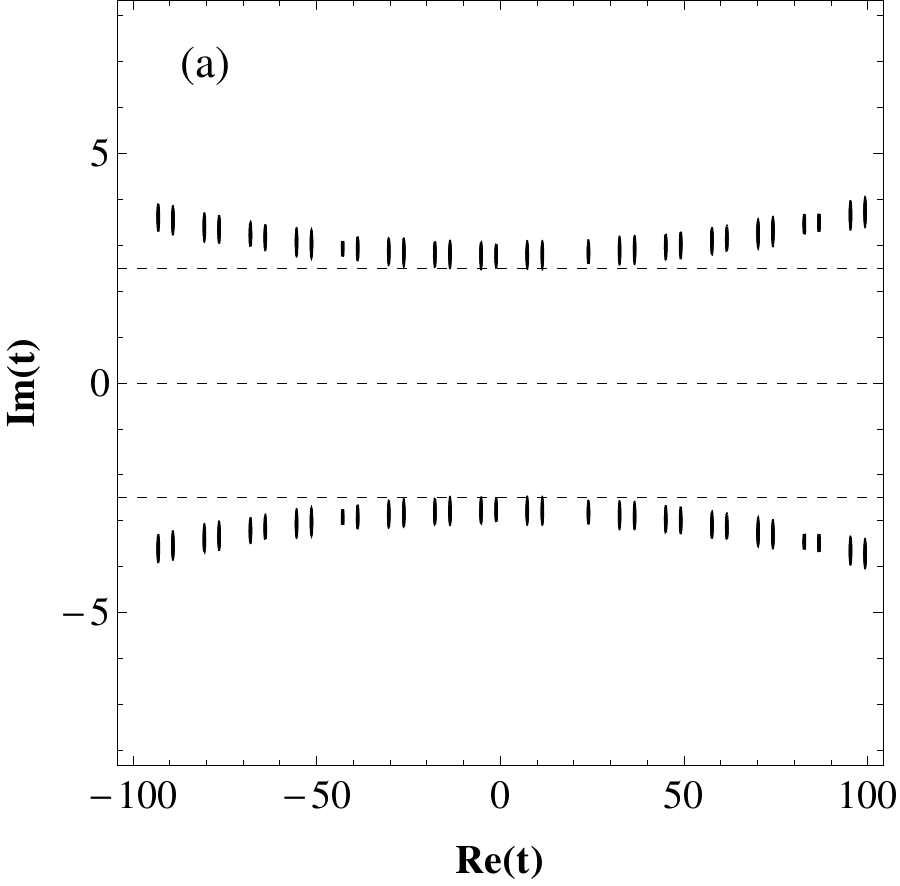}}
\hspace{7mm}
\subfigure{\includegraphics[width=6.8cm]{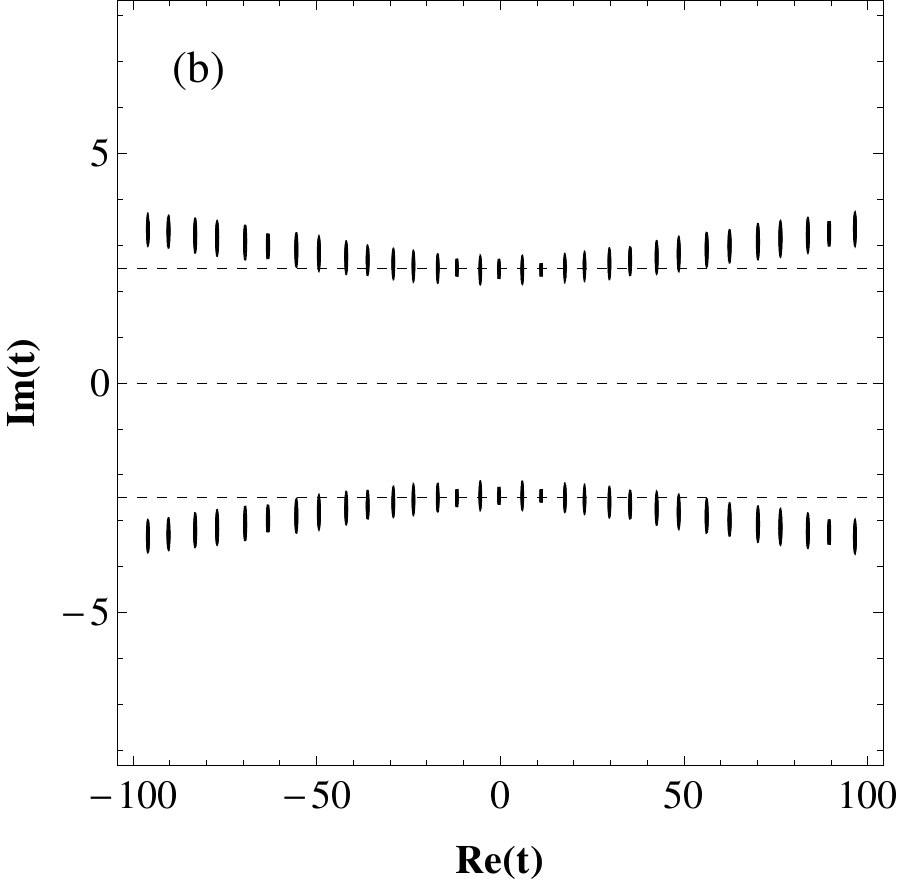}}
\hspace{7mm}
\subfigure{\includegraphics[width=6.8cm]{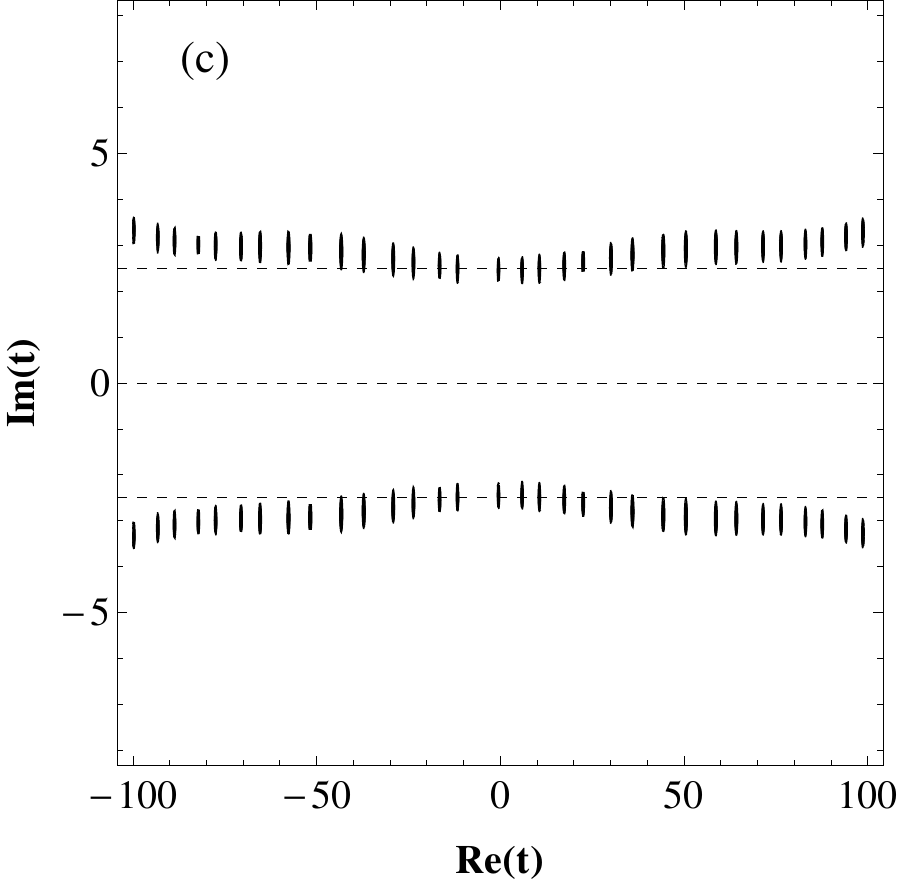}}
\hspace{7mm}
\subfigure{\includegraphics[width=6.8cm]{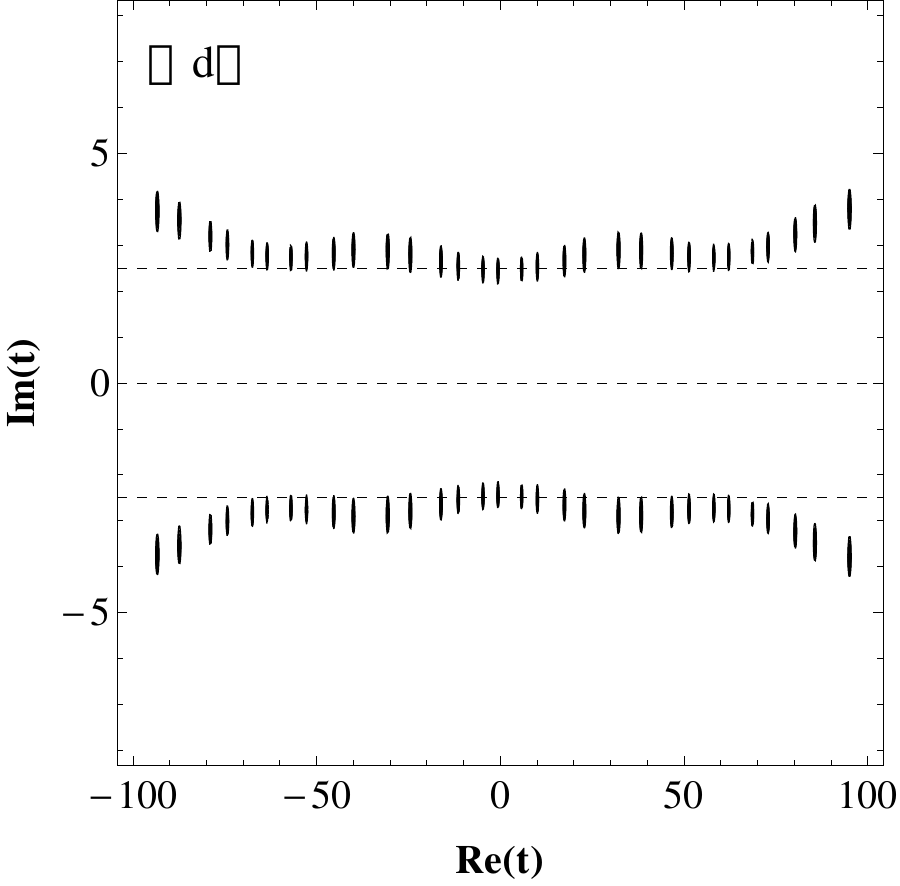}}
\setlength{\belowcaptionskip}{-0.3cm}
\caption{(color online). Turning points where $\omega_{\bf p}(t)=0$ by the contour plots of $|\omega_{\bf p}(t)|^2$ in the complex $t$ plane. Note that the plotting is got just for the time-dependent field where the spatial-dependent part is dropped, which can be regarded as the limit of large $\lambda$. Four sets of the momentum value are chosen, $p_{x}=0.664$, $0.190$, $0.288$ and $0.405$, respectively, corresponding to the maximum peak of the momentum spectrum as in the cases of $\lambda=1000$ in Fig.~\ref{1}(a) when $b=0$ and Figs.~\ref{2}[(a),(b) and (d)] when $b=1$, for the plotting from (a) to (d).}
\label{9}
\end{figure}

\subsection{Nonlinear effects on the reduced particle number}

In order to see why the reduced particle number appears obvious nonlinear effects on spatial scale when $\omega_{m}=0.07$ and $\omega_{m}=0.1$ in Fig.~\ref{3}(b), we employ the view point from the action of worldline instanton and the corresponding position distribution to make some discussions on this results, respectively.

It is known from Ref. \cite{Linder:2015} that the pair production rate can be written as $P_{e^{+}e^{-}}\sim e^{-S}$, where $S$ denotes the action of worldline instanton. In the time-dependent oscillation electric field $E(t)=E_0g(\omega t)$, the Keldysh adiabatic parameter is $\gamma_{\omega}=m\omega/eE_0$, where $\omega$ denotes the frequency of electric field. The well known result is that the corresponding $S_{\omega}$ decreases monotonically with $\gamma_{\omega}$ \cite{Gies:2005bz}, therefore, it leads to a monotonically increasing of the pair production rate $P_{e^{+}e^{-}}$. In contrary, if the field is only space-dependent oscillation as $E(x)=E_0 f(kx)$, the corresponding adiabatic parameter is expressed as $\gamma_{k}=mk/eE$, where $k$ represents wave number, and $S_{k}$ increases monotonically with $\gamma_{k}$ \cite{Gies:2005bz}. Therefore, the $P_{e^{+}e^{-}}$ decreases monotonically. Now it is noted that our electric field model, Eq.\eqref{FieldMode}, is composed by the space and time dependence, therefore, it should exhibit a complicated behavior due to the competition between the pair number increasing by the larger frequency and the pair number decreasing by the higher wave number.

Indeed, in Fig.~\ref{3}(b), it can be seen that when $\omega_{m}=0.07$ or $\omega_{m}=0.1$, the particle number appears an obvious nonlinear variation with spatial scale $\lambda$, i.e., there is a transition point at $\lambda=7$. It is noticed that the left and right sides of the transition point correspond to $\lambda=5$ and $9$, respectively. According to the description of the space-dependent field above, the particle number should enhance monotonously as spatial scale $\lambda$ increases (the corresponding $k$ decreases), but it increases nonlinearly, which is caused by the competitive relationship between the effect of the temporal part of the field on the particle number and the effect of the spatial part on the particle number. In the case of $\omega_{m}=0.07$, when the spatial scale $\lambda$ changes from $5$ to $7$, the effect of the temporal part of the field should be greater than that of the spatial part, which leads to a net effect of the increasing of the particle number. However, when the spatial scale varies from $7$ to $9$, the situation is vice versa, i.e., the effect of the temporal part of the field is smaller than the spatial part, resulting in a net effect of the decreasing of the particle number. Therefore, the particle number presents an obvious nonlinear variation with spatial scale. The discussion about the case of $\omega_{m}=0.1$ is similar to that of $\omega_{m}=0.07$.

On the other hand, the result can be understood by the position distributions for $\omega_{m}=0.07$ and $0.1$ in Fig.~\ref{10}. It can be seen from the Fig.~\ref{10}(a), for $\omega_{m}=0.07$, the position distribution presents the largest peak and the broadest distribution at $\lambda=7$ compared with the case of $\lambda=5$ and $9$, which indicates that the more particles can be created in the position space. The phenomenon and discussion of Fig.~\ref{10}(b) is similar to the case of Fig.~\ref{10}(a). Moreover, for $\omega_{m}=0.07$, we can also read the maximum peak values of the momentum spectra at $\lambda=5$, $7$ and $9$, as shown in the upper part of Table~\ref{Table 2}. For $\lambda=7$, the corresponding maximum values are $0.0277$ and $0.0292$, which are larger than the case of $\lambda=5$ and $9$. Therefore, the corresponding reduced particle number is also the largest, i.e., $\bar{N}=0.0129$. When $\omega_{m}=0.1$, the maximum peak values of the momentum spectra at $\lambda=5$, $7$ and $9$ are shown in the lower part of Table~\ref{Table 2}, where the result is similar to the case of $\omega_{m}=0.07$.

\begin{figure}[H]%\suppressfloats
\begin{center}
\includegraphics[width=\textwidth]{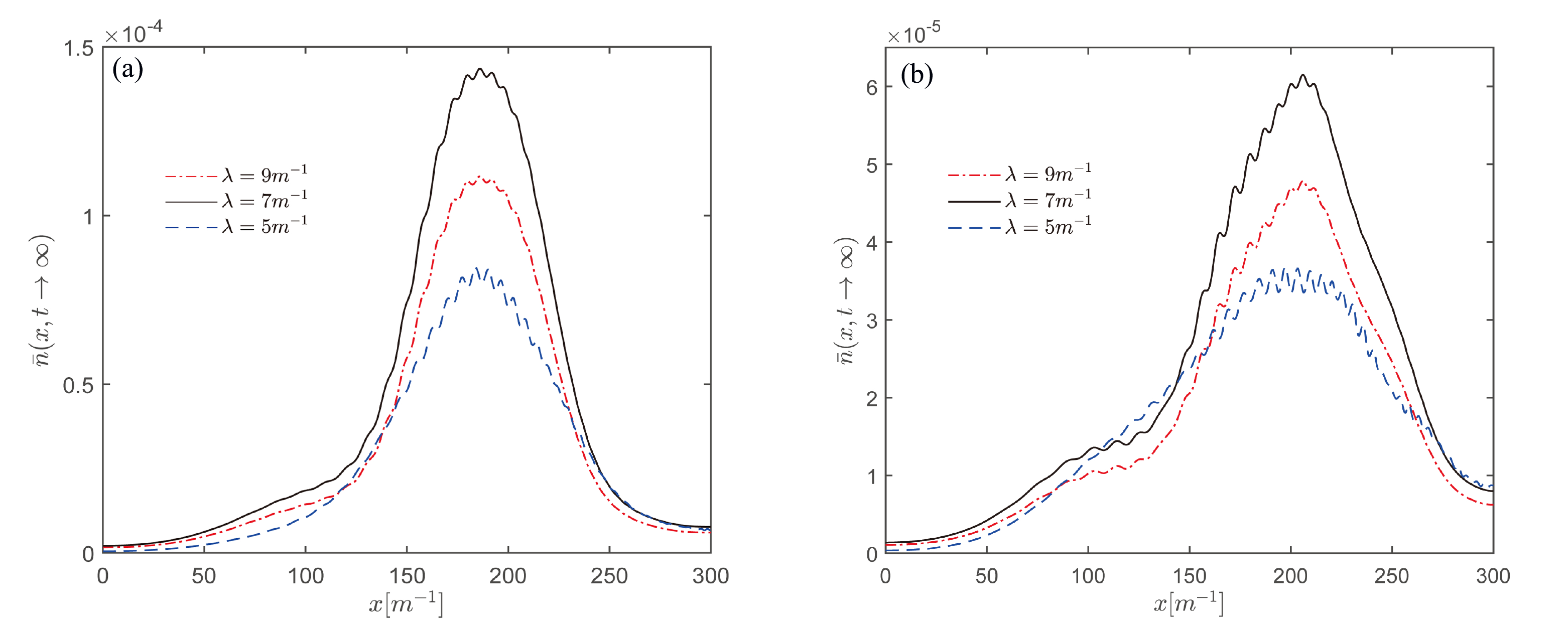}
\end{center}
\setlength{\abovecaptionskip}{-0.3cm}
\setlength{\belowcaptionskip}{-0.3cm}
\caption{(color online). The position distribution for various spatial scales in high frequency field with different modulated frequency values. They are $\omega_{m}=0.07$ for (a) and $\omega_{m}=0.1$ for (b), respectively. Other field parameters are the same as in Fig.~\ref{3}(b).}
\label{10}
\end{figure}

\begin{table}[H]
\caption{The maximum peak values of the momentum spectrum and the reduced particle number for various spatial scales in high frequency field with different modulated frequency parameters. The modulated frequency is $\omega_{m}=0.07$ for the upper part of table and the modulated frequency is $\omega_{m}=0.1$ for the lower part of table. Other field parameters are $\epsilon=0.3E_{\mathrm{cr}}$, $\omega=0.5$, $b=1$ and $\tau=100$.}
\centering
\begin{ruledtabular}
\begin{tabular}{cccc}
$\lambda$ &$({p}^{*}, \bar{n}({p}^{*},t\rightarrow\infty)_{max})_{left}$    &$({p}^{*}, \bar{n}({p}^{*},t\rightarrow\infty)_{max})_{right}$  &$\bar{N}$\\
\hline
$5m^{-1}$    &$(-0.449, 0.0145)$       &$(0.488, 0.0142)$              &$0.00816$\\
$7m^{-1}$    &$(-0.474, 0.0277)$       &$(0.483, 0.0292)$              &$0.0129$\\
$9m^{-1}$    &$(-0.479, 0.0215)$       &$(0.479, 0.0229)$              &$0.0120$\\
\hline
$\lambda$ &$({p}^{*}, \bar{n}({p}^{*},t\rightarrow\infty)_{max})_{left}$    &$({p}^{*}, \bar{n}({p}^{*},t\rightarrow\infty)_{max})_{right}$  &$\bar{N}$\\
\hline
$5m^{-1}$    &$(-0.376, 0.00681)$       &$(0.508, 0.00545)$            &$0.00503$\\
$7m^{-1}$    &$(-0.454, 0.0105)$        &$(0.547, 0.00983)$            &$0.00690$\\
$9m^{-1}$    &$(-0.459, 0.00814)$       &$(0.562, 0.00763)$            &$0.00536$\\
\end{tabular}
\end{ruledtabular}
\label{Table 2}
\end{table}

\section{CONCLUSION AND OUTLOOK}\label{conclusion}

In conclusion, with the DHW formalism, we have investigated the sinusoidal phase modulation on the momentum spectrum and the reduced particle number in both
high- and low-frequency inhomogeneous fields. The effect of spatial scale of the external field on the pair production is further examined. Furthermore, we give a qualitative discussion of some results that we obtained by using the semiclassical WKB approximation and the view point from the action of worldline instanton.

For high frequency field, the increasing of either modulated frequency or amplitude leads to significant interference effect on the momentum spectrum, but with modulated amplitude, the symmetry of the momentum spectrum is destroyed severely, while momentum spectrum shows a good symmetry with modulated frequency. The reduced particle number is enhanced significantly by the variation of modulation parameters for different spatial scales. At small spatial scales, the particle number is enhanced by more than one order of magnitude with modulated amplitude, while it is increased about $5$ times with modulated frequency. At large spatial scales, compared with the case without modulation, the particle number is increased by more than one order of magnitude for either large modulated amplitude or small modulated frequency. Moreover, two interesting features are revealed for the reduced particle number, the optimal modulation parameters are obtained and the same particle number can be got through different set of modulation parameters.

For low frequency field, the nonlinear effects lead to the more and more complicated momentum peaks with modulated amplitude. The reduced particle number is enhanced greatly with either modulated amplitude or frequency even at extremely small spatial scale. Specifically, for modulated amplitude, the particle number is increased about $3$ times, while it is enhanced about $9$ times for modulated frequency. Moreover, we examined the effect of spatial scale on the particle number. When modulation parameters are fixed, the reduced particle number increases rapidly at small spatial scales, while it tends to be a constant at large spatial scales. Meanwhile, the particle number presents an obvious nonlinear variation for the certain modulated frequencies at small spatial scale. Finally, we found that the momentum spectrum is more sensitive to modulated amplitude, while the reduced particle number is more sensitive to modulated frequency.

Our study indicates that the sinusoidal phase modulation plays a crucial role on $e^{-}e^{+}$ pair production in spatially inhomogeneous electric fields, meanwhile, it also provides the theoretical basis of broader parameter ranges for future experiments. In this paper, we have only considered the influence of modulation parameters at various spatial scales on the pair production. For the further investigation, the carrier phase effect and dynamically assisted effect, which are caused by the combined field with sinusoidal phase modulation, should be considered on the $e^{-}e^{+}$ pair production. Furthermore, provided that the computational resources are sufficient, one can consider the $e^{-}e^{+}$ pair production in multidimensional spatially inhomogeneous external fields with the sinusoidal phase modulation.

\begin{acknowledgments}
\noindent
We thank M. Ababekri and M. A. Bake for fruitful discussions and their help with the numerical calculation. We are also grateful to Li Wang for critical reading of the manuscript. This work was supported by the National Natural Science Foundation of China (NSFC) under Grant No.\ 11875007 and No.\ 11935008. The computation was carried out at the HSCC of the Beijing Normal University.
\end{acknowledgments}

\end{document}